\pdfoutput=1
\documentclass{JINST}
\let\ifpdf\relax

\title{Three-Dimensional Triplet Tracking for LHC and Future High Rate Experiments}

\author{Andr\'e Sch\"oning \\
Physikalisches Institut, Universit\"at Heidelberg \\
  Im Neuenheimer Feld 226, 69120 Heidelberg, Germany \\
  E-mail: \email{schoning@physi.uni-heidelberg.de}}

\abstract{The hit combinatorial problem is a main 
challenge for track reconstruction and triggering 
at high rate experiments.
At hadron colliders the dominant fraction of hits is due to low momentum
tracks for which multiple scattering (MS) effects dominate the hit resolution.
MS is also the dominating source for hit confusion and track uncertainties in
low energy precision experiments.
In all such environments, where MS dominates, track reconstruction and fitting can
be largely simplified by using three-dimensional (3D) hit-triplets as provided by
pixel detectors.
This simplification is possible since track uncertainties are solely determined by MS if
high precision spatial information is provided. 
Fitting of hit-triplets is especially simple for tracking detectors
in solenoidal magnetic fields.

The over-constrained 3D-triplet method 
provides a complete set of track parameters and is robust against
fake hit combinations.
Full tracks can be reconstructed step-wise by connecting 
hit triplet combinations from different layers, 
thus heavily reducing the combinatorial 
problem and accelerating track linking. 

The triplet method is ideally suited for pixel detectors where hits can be
treated as 3D-space points. 
With the advent of relatively cheap and industrially available 
CMOS-sensors the construction of highly granular full scale pixel tracking 
detectors seems to be possible also for experiments at LHC or 
future high energy (hadron) colliders.
In this paper tracking performance studies for full-scale pixel detectors, 
including their
optimisation for 3D-triplet tracking, are presented. 
The results obtained for different types of tracker geometries and different
reconstruction methods are compared.
The potential of reducing the number of tracking layers and - along with that -
the material budget using this new tracking concept is discussed.
The possibility of using
3D-triplet tracking for triggering and fast online reconstruction is 
highlighted.}

\keywords{LHC; Track Reconstruction; Track Trigger}

\begin{document}
\section{Introduction}
Recent advances in semiconductor technology make it possible to produce pixel sensors
at moderate costs. 
In particular Monolithic Active Pixel Sensors (MAPS) are very
attractive as they allow to integrate the readout logic in the sensors
itself, thus simplifying the detector fabrication (no bump bonding) and 
reducing material and power consumption. 
A particularly interesting variant are High-Voltage MAPS
(HV-MAPS)~\cite{ref:peric1,ref:peric2}  as they
collect the charge by drift and not by diffusion as standard MAPS sensors do.
This feature allows to use CMOS sensors also in high rate experiments like LHC. 
Recent tests of HV-MAPS prototype sensors~\cite{ref:HVMAPS_Results} show
promising results and make people think about the idea to construct 
silicon-pixel only tracking detectors for future high rate and precision
experiments. 
The dream to use an all-silicon pixel detector will soon come to reality in the Mu3e
experiment~\cite{ref:Mu3e_RP,ref:Mu3e_WIT14},
which aims to search for the lepton flavor violating $\mu \rightarrow eee$ decays
with a sensitivity of better than $10^{-16}$. The Mu3e experiment will mainly consist
of a four-layer pixel detector
with an active surface of about 1.5~m$^2$ and almost 300$\,$000 pixels.
However scaling of such a big pixel detector to even larger sizes, for
example LHC experiments, is not trivial at all. 
The main obstacles are the power dissipation of the active sensors
and the small reticle sizes of the CMOS technology. 
On the other hand HV-MAPS have many advantages as they allow for a  much
simpler readout architecture (hit digitisation on sensor), reduced material budget 
(thinned sensors and no additional readout chips) and simplified detector 
design (no strip stereo layers, fewer layers, etc.). 
\begin{figure}[thb]
\label{fig:1} 
\center
\includegraphics[width=0.6\textwidth] {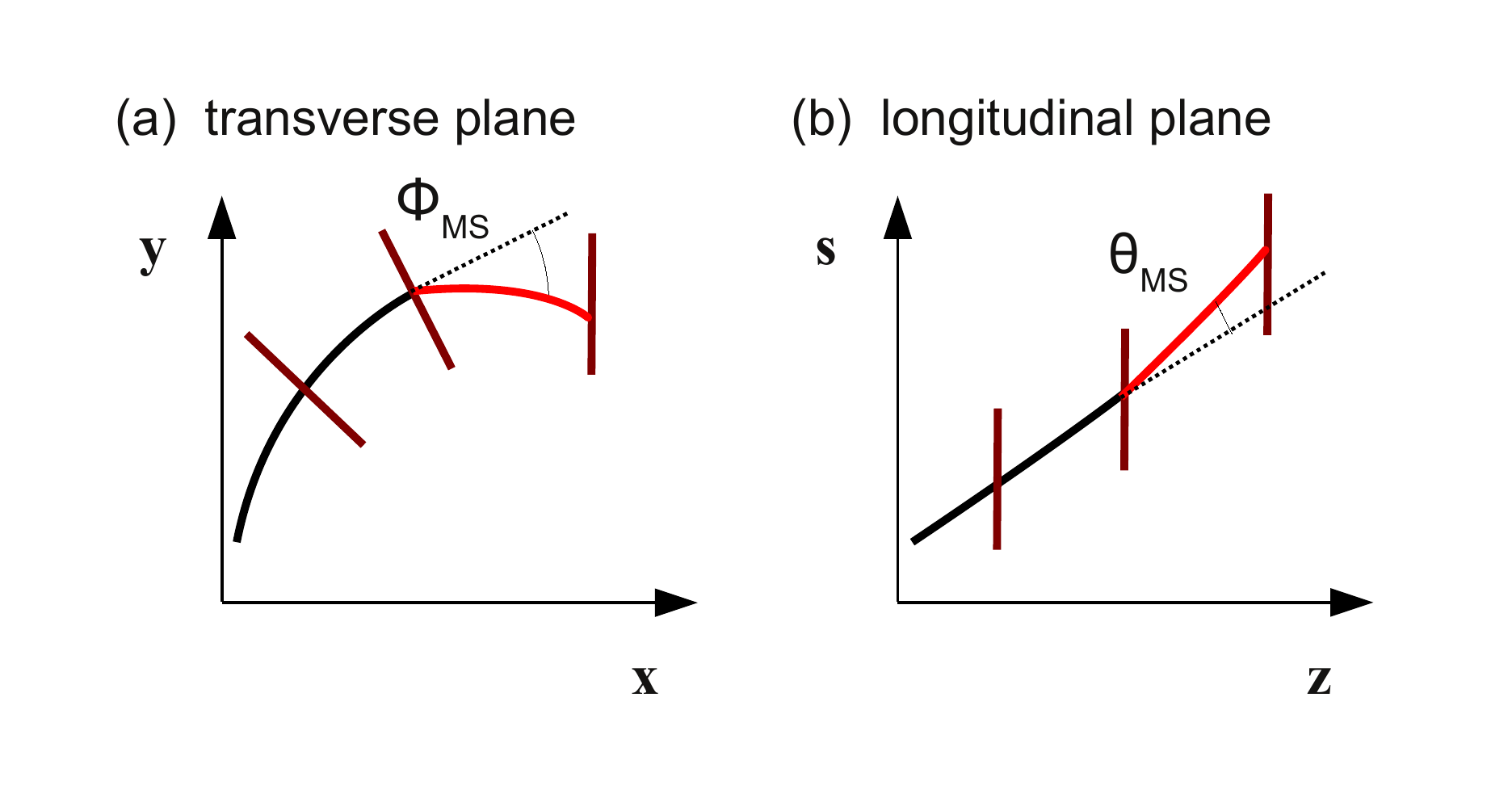}
\vspace{-0.5cm}
\caption{
Sketch of multiple scattering at the middle triplet layer in the
(a)~transverse and (b)~longitudinal plane with the scattering angles denoted as
$\phi_{\rm MS}$ and $\theta_{\rm MS}$, respectively. 
}
\end{figure}

A big advantage of pixel detectors is that only three sensor layers are needed
to provide full tracking information. 
Because of the higher redundancy, 
the number of detector layers can be significantly reduced in a
pixel-only design compared to a design based on silicon strips.
Thus, the  pixel related increase of the power
consumption is expected to be fully compensated by reducing the total sensor surface.
With the same argument 
the total readout bandwidth of a pixel-only tracking detector is expected 
to be even smaller than for a strip-based detector.
A further reduction of the bandwidth is expected from the much smaller
hit-cluster size of thinned pixel sensors as provided by the HV-MAPS
technology.

In the following we first discuss
the basic properties of hit-triplets including multiple scattering
effects.
Thereafter we introduce different pixel-detector design concepts with
emphasis on the tracking resolution and the hit-combinatorial problem.
We compare the ``vector tracking'' and ``triplet tracking''
concept for trigger applications and discuss the advantages of using triplet-layers
in the tracking detector design in terms of resolution and triggering.

\section{Hit triplets}
\label{sec:triplet}

Here and in the following we assume that tracks are measured in a homogeneous
magnetic field where trajectories of charged particles are described by 
helices. 
Seven parameters
are needed to describe a helix from a  start to an end-point\footnote{The parameters are: the start point (3), the direction (2), the
curvature (1) and the length of the helix (1).}.
Therefore, in order to reconstruct the track parameters at least three space
points are
needed which have to be measured. 
If solid state tracking detectors are used
a complication arises from MS.
Because of deflection at the middle space point (detector layer) two helices are needed to describe the
trajectory, see figure~\ref{fig:1}. 
The connected helices are then described by 
$2 \times 7 - 3 =11$ parameters. Only ten parameters are needed if the
momentum, and correspondingly the 3D track radius (inverse of
the 3D curvature $R_{\rm 3D} = 1/\kappa_{\rm 3D}$), do not
change\footnote{This is typically a very good assumption for thin silicon
sensors.}.
Two of the ten parameters are the multiple scattering angles at the middle
layer in the transverse plane
$\phi_{\rm MS}$ and longitudinal plane $\theta_{\rm MS}$.\footnote{
The longitudinal plane is defined as $s$-$z$ plane with $s$ being the 3D-track
  length and $z$ the direction parallel to the magnetic field.}.
With three hit coordinates, however, only nine parameters can be determined,
thus one parameter remains undefined. 
%The situation does not change if we add more hits. Also 
Note that hit uncertainties have been neglected so far.

\begin{figure}[tb]
\label{fig:2} 
\center
\includegraphics[width=0.75\textwidth] {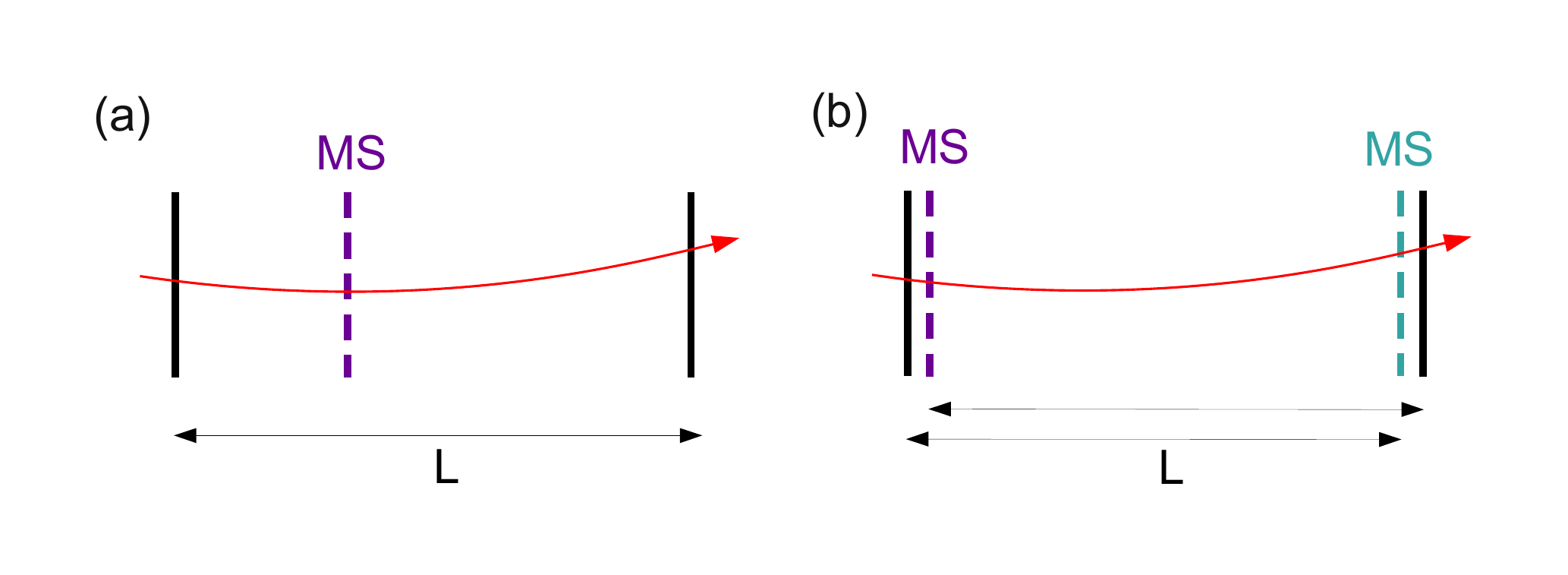}
\vspace{-0.5cm}
\caption{
(a) Sketch of a three-layer and (b) 
four-layer geometry for (optimal) track reconstruction if multiple scattering
dominates.
The relevant MS-layers determining the momentum resolution are indicated.}
\end{figure}

A solution to this problem comes from scattering theory which says that the
average scattering angle vanishes. The RMS of the scattering angle 
distribution is given by $\Theta_{RMS} \propto \sqrt{X/X_0}/p$ with 
$p$ being the particle momentum and $X/X_0$ the material thickness in units
 of the radiation length~$X_0$.
In the following we use the Highland formula as given by the Particle Data
Group~\cite{ref:PDG}. 
By adding two constraints from the expected scattering distribution 
and assuming energy conservation the two helices
can now be fitted.
The general solution of this triplet fit, which can be found by a 
linearisation ansatz~\cite{ref:triplet},
can be directly calculated from the coordinates of the three hit positions.

Now we turn to the question which detector design, if MS dominates, provides the best
track (momentum) resolution for a given distance $L$ between the first and
last hit point, see figure~\ref{fig:2}~(a). 
For the relative momentum resolution we obtain:
\begin{equation}
\label{eq:2.1}
\frac{\sigma_{R_{\rm 3D}}}{R_{\rm 3D}} = \frac{\sigma_{p}}{p} = 
\frac{2 b}{B L} \quad 
\end{equation}
with $b$ being an effective  scattering parameter describing the thickness of
material at the middle hit position and $B$ the magnetic field strength.\footnote{Here we assume no material between the
  sensor layers.} 
Note that spatial the resolution is independent of the relative position of the middle
layer with respect to the first and last layer. 
Using (\ref{eq:2.1}), it is easy to prove that for given $L$ and dominating MS uncertainties
the best track measurement is obtained for a 
configuration with four layers as shown in figure~\ref{fig:2}~(b) 
where two closely stacked pairs of
layers are used. For this configuration one obtains:
\begin{equation}
\frac{\sigma_{R_{\rm 3D}}}{R_{\rm 3D}} = \frac{\sigma_{p}}{p} = 
\frac{\sqrt{2} b}{B L} \quad ,
\end{equation}
providing a factor $1/\sqrt{2}$ reduction in resolution compared to any
three-layer configuration. This reduction comes here from the 
combination of the two triplet measurements which are statistically independent
(two different scattering layers).
Also note that the relative momentum resolution is momentum independent as
long as multiple scattering dominates and non-linear geometrical effects from
bending can be neglected.

\begin{figure}[thb]
\center
\includegraphics[width=0.65\textwidth] {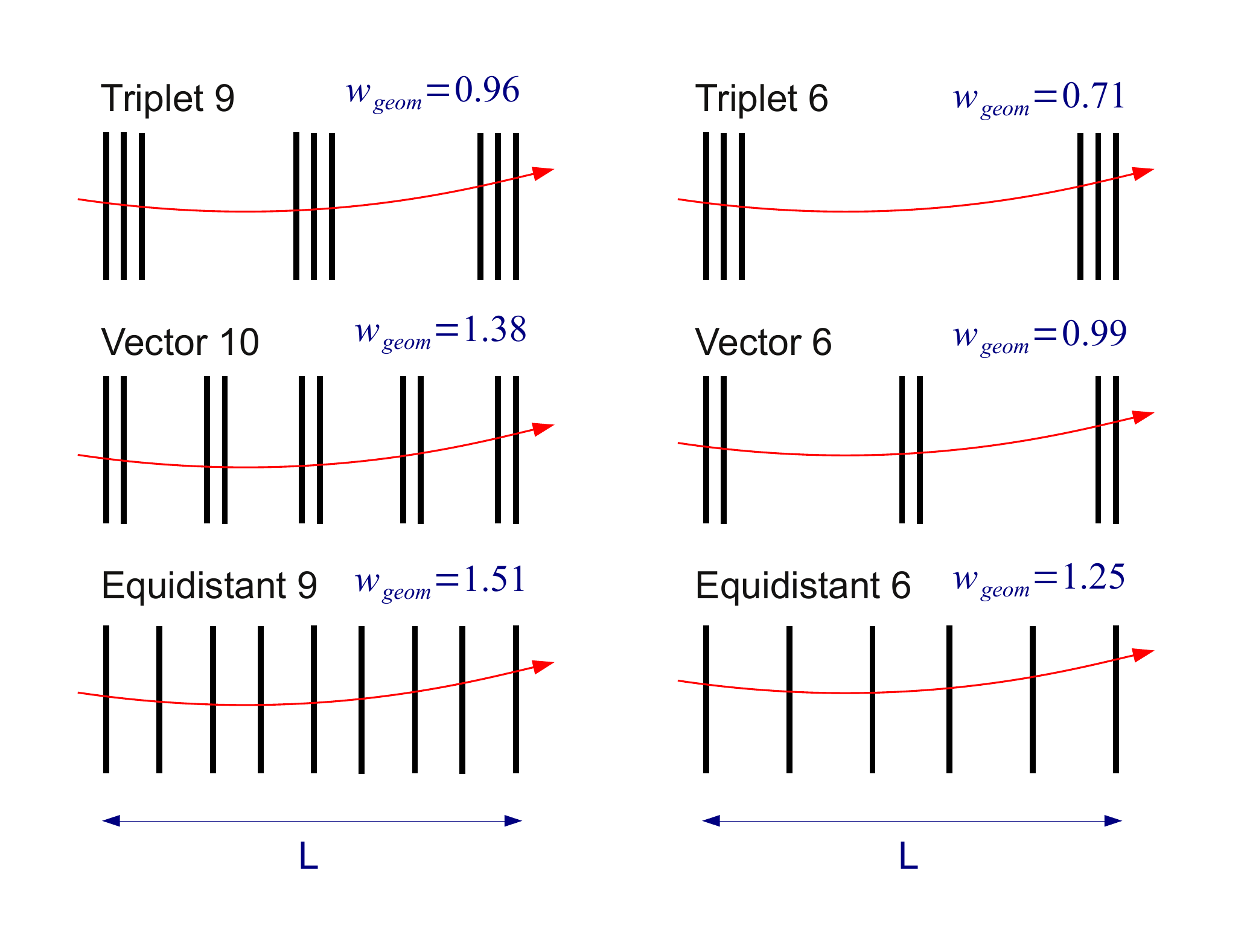}
\vspace{-0.5cm}
\caption{
\label{fig:3} Different types of detector geometries used for simulation
studies. The multiple scattering geometry parameter $w_{\rm geom}$ 
determines the relative momentum resolution if MS dominates.
}
\end{figure}

\section{Optimal detector geometry and multiple scattering}
Above we have discussed optimal geometries for spectrometers considering MS.
But why should we discuss MS for
a search machine like LHC where mostly high momentum particles 
exhibiting low MS are used for triggering and new physics searches?
The answer is simple; most particles $\approx 99\%$ produced at LHC have small transverse
momentum $p_T \lesssim 1$~GeV, and
%. 
%Most of them originate from pile up and underlying events.
%They also originate from high momentum jets and are important for measurements.
%Furthermore,
low momentum tracks are responsible for the 
hit combinatorial problem which is the main complication for (fast) track
reconstruction especially at high luminosities (pile-up). 
Therefore, MS has to be considered for an optimised tracker layout.

In figure~\ref{fig:3} different types of detector geometries are shown
which are studied in more detail in the following sections. 
For the simulation we assume an axial-symmetric tracking detector. 
The layer planes in figure~\ref{fig:3} refer to the relative radial positions.
For each of the different geometries we can parameterise the relative
momentum resolution if MS dominates as:
\begin{equation}
\frac{\sigma_{R_{\rm 3D}}}{R_{\rm 3D}} = \frac{\sigma_{p}}{p} = 
\omega_{\rm geom} \; \frac{{2} b}{B L} \quad , 
\end{equation}
where a generalised geometry factor $\omega_{\rm geom}$ is introduced.
%In figure ~\ref{fig:3} 
We consider three classes of detector types: geometries
with equidistant layer spacing, geometries with closely stacked doublet layers
and geometries with closely stacked triplet layers.
The design with doublet layers are also referred to as ``vector'' as they allow
to measure the direction of the particle at each doublet layer. 
For all geometries the $w_{\rm geom}$ parameters,  as calculated from scattering
theory, are given. They are calculated for small curvatures 
and infinitesimally small distances between stacked
layers.
Theoretically, the ``Triplet6'' geometry provides the best resolution. This can
be easily understood as it 
approaches the optimal design of figure~\ref{fig:2}~(b) for infinitesimal
stacking. 
It is interesting to note that equidistant designs have the worst resolution
if MS dominates and that the relative momentum resolution scales as
$\approx \sqrt{N_{\rm layer}}$ if the  number of layers  is large.

\begin{figure}[th]
\setlength{\unitlength}{1.0cm}
\center
\includegraphics[width=0.48\textwidth] {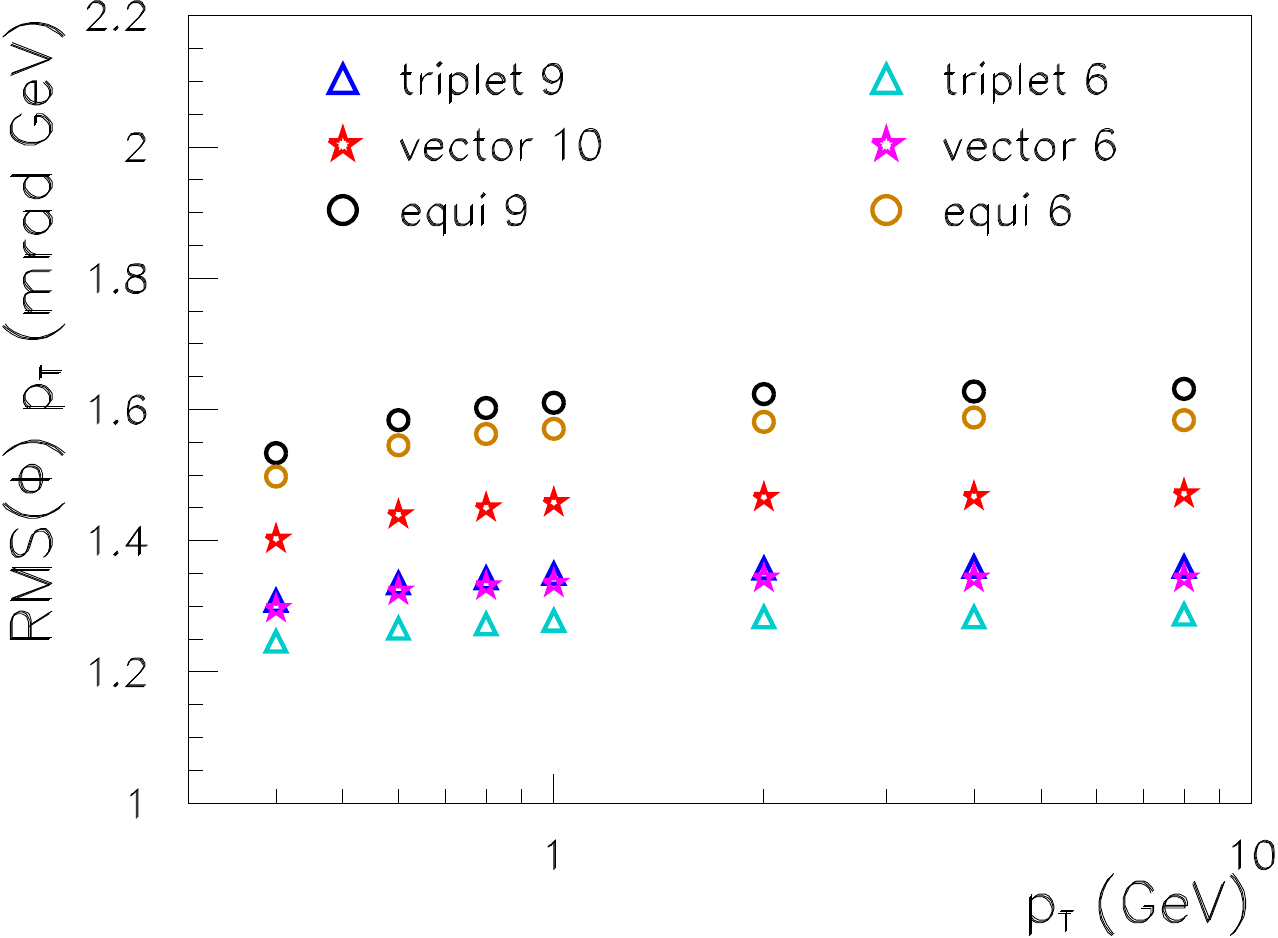}
\put(-6.9,5.7) {(a)}
~~~
\includegraphics[width=0.48\textwidth] {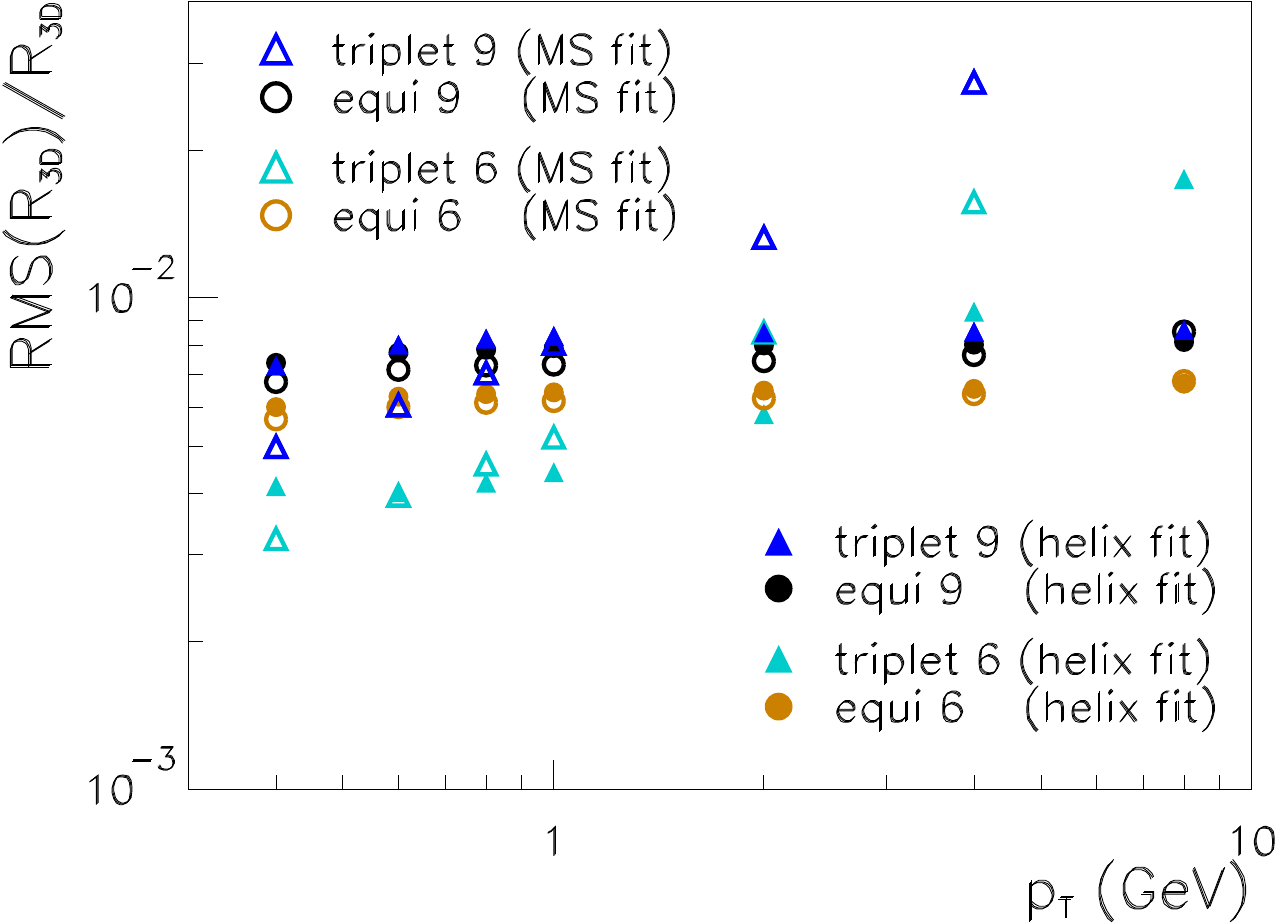}
\put(-6.9,5.7) {(b)}
\caption{
\label{fig:4} (a) Simulated RMS uncertainty of the $p_T$-normalised azimuthal track angle
$\phi_{\rm vtx}$ at
the origin (vertex) as function of $p_T$ for various detector
geometries. Only multiple scattering  is 
simulated and fitted using $1\%~X_0$ per layer.
(b) RMS of the relative momentum resolution as function of $p_T$ for
various detector geometries. An additional hit position uncertainty of $12~\mu$m
is simulated. The results are shown for two different
fits: multiple scattering fit ({\sl open symbols}) and helix fit ({\sl full markers}).
The detector geometries are defined in figure~\protect\ref{fig:3}.
}
\end{figure}

\section{Track parameter resolution study}
\label{sec:simulation}

A Monte Carlo simulation is used to study for the different geometries
the resolution of the reconstructed 
track parameters: the azimuthal angle
$\phi_{\rm vtx}$, the polar angle $\theta_{\rm vtx}$, the track radius $R_{\rm
  3D}$, and the
distance of closest approach to the origin in the transverse plane $D_0$ 
and in the longitudinal plane $Z_0$.
For the simulation we leave the inner detector region un-instrumented and place the first layer at a radius of $r=20$~cm and 
the last layer at  $r=100$~cm. 
The distance between stacked layers is chosen to be $\Delta r=1$~cm for both
vector and triplet layers. Minimum bias events are generated and the track
parameters resolutions are studied in the central detector region $-1.5< \eta < 1.5$~\footnote{
$\eta$ is the pseudorapidy of the track.
}.
We use a magnetic field of $B=2$~T and set the material thickness per layer to
$1\%$ radiation length, resulting in a material parameter of $b \approx 0.005$~Tm.

The largest differences between the studied geometries are seen in the $\phi_{\rm vtx}$
resolution and in the momentum resolution. 
In figure~\ref{fig:4}~(a) the RMS of the $p_T$-normalised $\phi_{\rm vtx}$ resolution 
%defined as $RMS(\phi_{\rm vtx})^2 = \overline{(\phi_{\rm vtx, fit}-\phi_{\rm vtx, gen})^2}$
is shown as function of the track $p_T$  using the MS-fit. 
Note that the fit consists of many helix parameterisations. 
In general, designs
with large gaps between the layers provide
a better resolution.
The triplet tracking design with six layers has about $20\%$ better
$\phi$ resolution than the equidistant 9-layer design. 

\begin{figure}[th]
\label{fig:5} 
\setlength{\unitlength}{1.0cm}
\center
\includegraphics[width=0.45\textwidth] {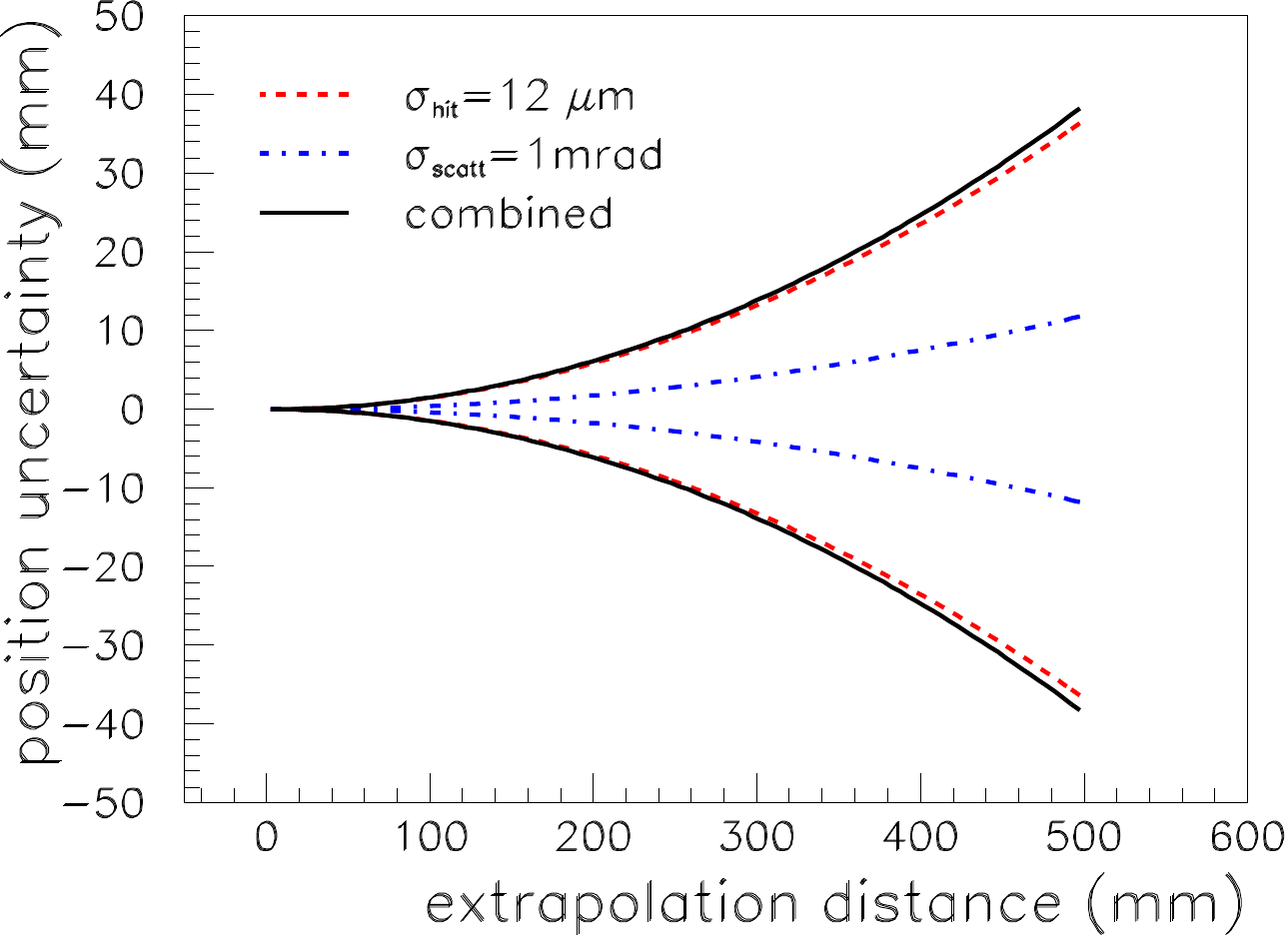}
\put(-6.5,5.2) {(a) transverse}
~
\includegraphics[width=0.45\textwidth] {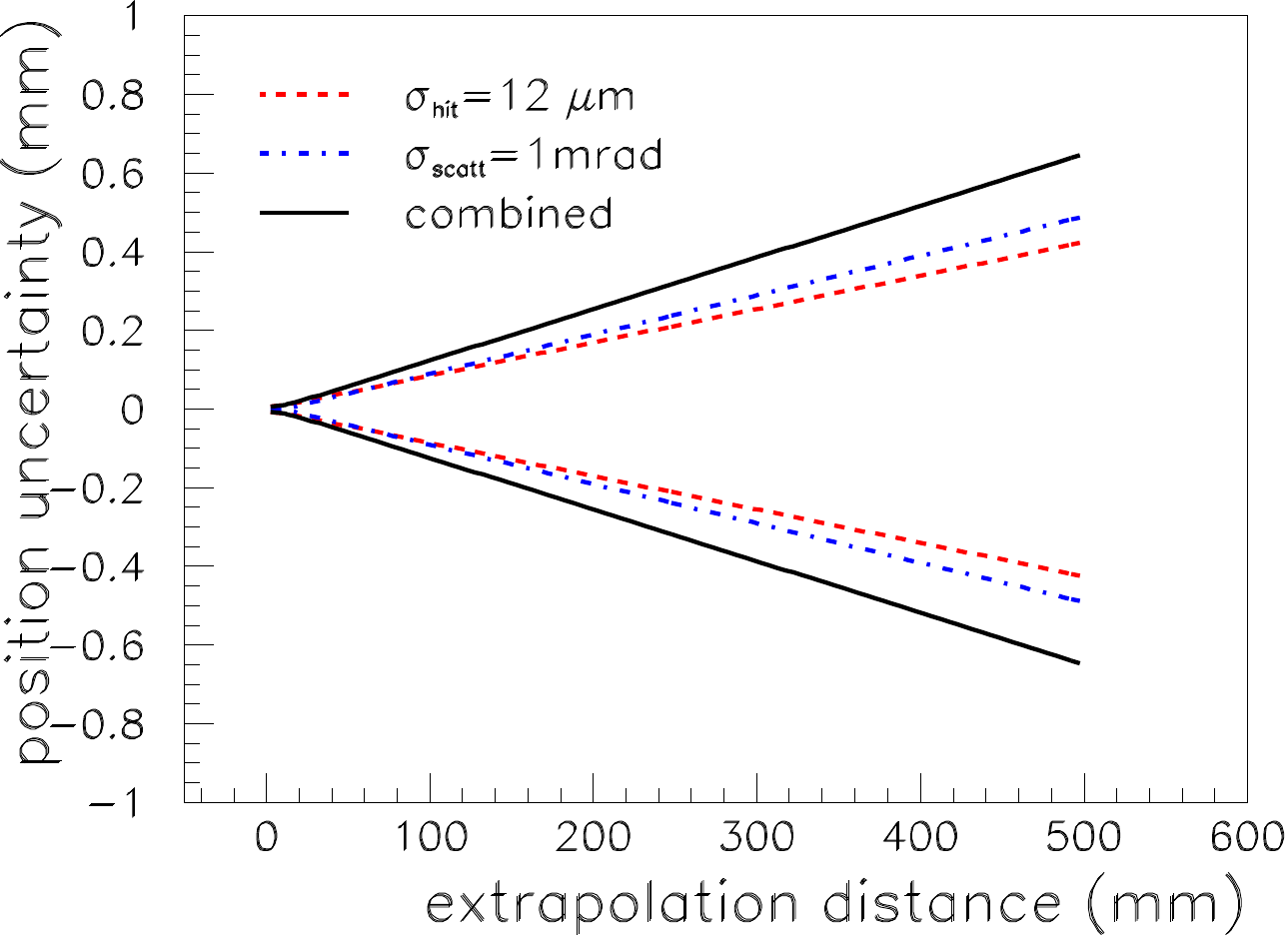}
\put(-6.5,5.2) {(b) longitudinal}
\caption{
Calculated $1\sigma$ envelope of the extrapolated track
with parameters reconstructed from a hit triplet in the
 (a)~transverse and (b)~longitudinal plane.
 The distance between the triplet hits is $1$~cm, the hit position uncertainty is
 $\sigma_{\rm hit}=12~\mu$m and the multiple scattering uncertainty is
 $\sigma_{\rm scatt}=1$~mrad. The contributions from the hit position
 uncertainty ({\sl red dashed}) and from scattering ({\sl blue dashed-dotted})
are shown.
}
\end{figure}

In figure~\ref{fig:4}~(b) the RMS of the relative momentum resolution is shown
for two different fits: the MS-fit and a global helix fit~\cite{ref:karimaki}
where all hits are fitted by an unique helix parameterisation.
In addition to MS, hit position uncertainties are also simulated here using
a hit resolution of $\sigma_{hit}=12~{\mu}$m (corresponding to a pixel size of
$40 \times 40~\mu{\rm m}^2$).
At low momentum, where MS effects dominate, the MS-fit gives the best momentum
resolution and the triplet layer designs are superior compared to
the equidistant designs.
With increasing $p_T$ the resolution obtained by the global helix fit is rather constant
whereas the resolution of the MS-fit gets worse, in general.
A cross over point can be defined where the results
of the MS-fit and the global helix fit are comparable.
For the triplet designs and vector designs (not shown) the cross over point is
in the $p_T \approx 1$~GeV region whereas for the equidistant designs the cross
over point is at $p_T \approx 10$~GeV. 
This can be easily understood; for stacked layers with small layer spacings
the angular uncertainty of each track segment, $\sigma_{\rm
  hit}/\Delta r$, is much larger than for designs with equidistant spacing and large
gaps between the layers. 
Therefore, hit uncertainties in designs with stacked layers affect the momentum
resolution at much lower $p_T$-values.
For tracks with $p_T \gtrsim 1$~GeV both multiple scattering effects and hit
uncertainties have to be included for a proper track fit.
Such combined fits are performed by, for example, Kalman filters~\cite{ref:kalman} and General Broken Line
fits~\cite{ref:GBL}.
At very high momentum, $p_T \gtrsim 100$~GeV, (not
shown in the figures)  MS effects are negligible, and
the equidistant and stacked layer designs show a very similar
performance.
% with the resolution scaling as $1/N_{\rm layer}$ as long as pivot points are
%provided at middle radii (applies to all designs except ``triplet 6'').

To summarise, about $99\%$ of all particles at LHC are of low
momentum. 
Low $p_T$ tracks for which MS dominates can  be best measured with an optmised tracker design
with large gaps between sensor layers and a minimum number of
layers. These tracks are most efficiently
reconstructed using triplet fits.

\section{Track linking study}

Track linking is the most time expensive step in the track reconstruction. 
Usually, a track seed is needed which is then extrapolated to the next layers.
A seed can either be defined by a pair of hits (3D space points) plus an
additional constraint, for example that the particle originates from a
collision point or beamline, or by a hit triplet which
already defines a track, see section~\ref{sec:triplet}.
Hit triplets are especially easy to find in a pixel tracker with triplet
layers. For not too low transverse momenta all three hits of a triplet can be
roughly connected
by a straight line. The relatively low occupancy in pixel
detectors makes it very easy to find such triplet seeds. 
Next, the question arises how well we can extrapolated the trajectory from the
last layer of a seed to the next (more far distant) sensor layer?
Figure~\ref{fig:5} shows the $1\sigma$ envelopes of the extrapolated
trajectories from a hit triplet with a hit
distance of $\Delta r=1$~cm. 
\begin{figure}[tb]
\label{fig:6} 
\setlength{\unitlength}{1.0cm}
\center
\includegraphics[width=0.49\textwidth] {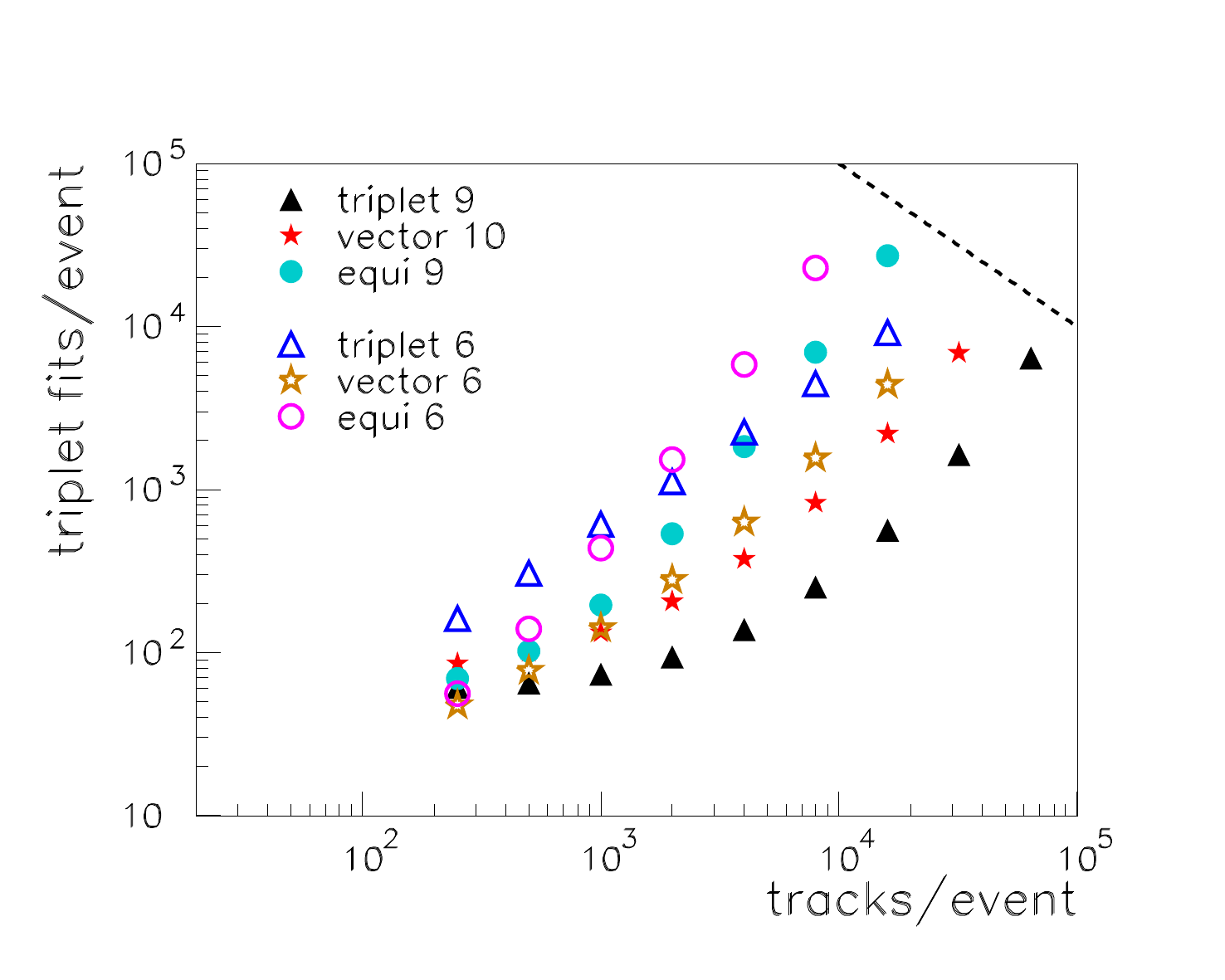}
%\hspace{-0.0cm}
\includegraphics[width=0.49\textwidth] {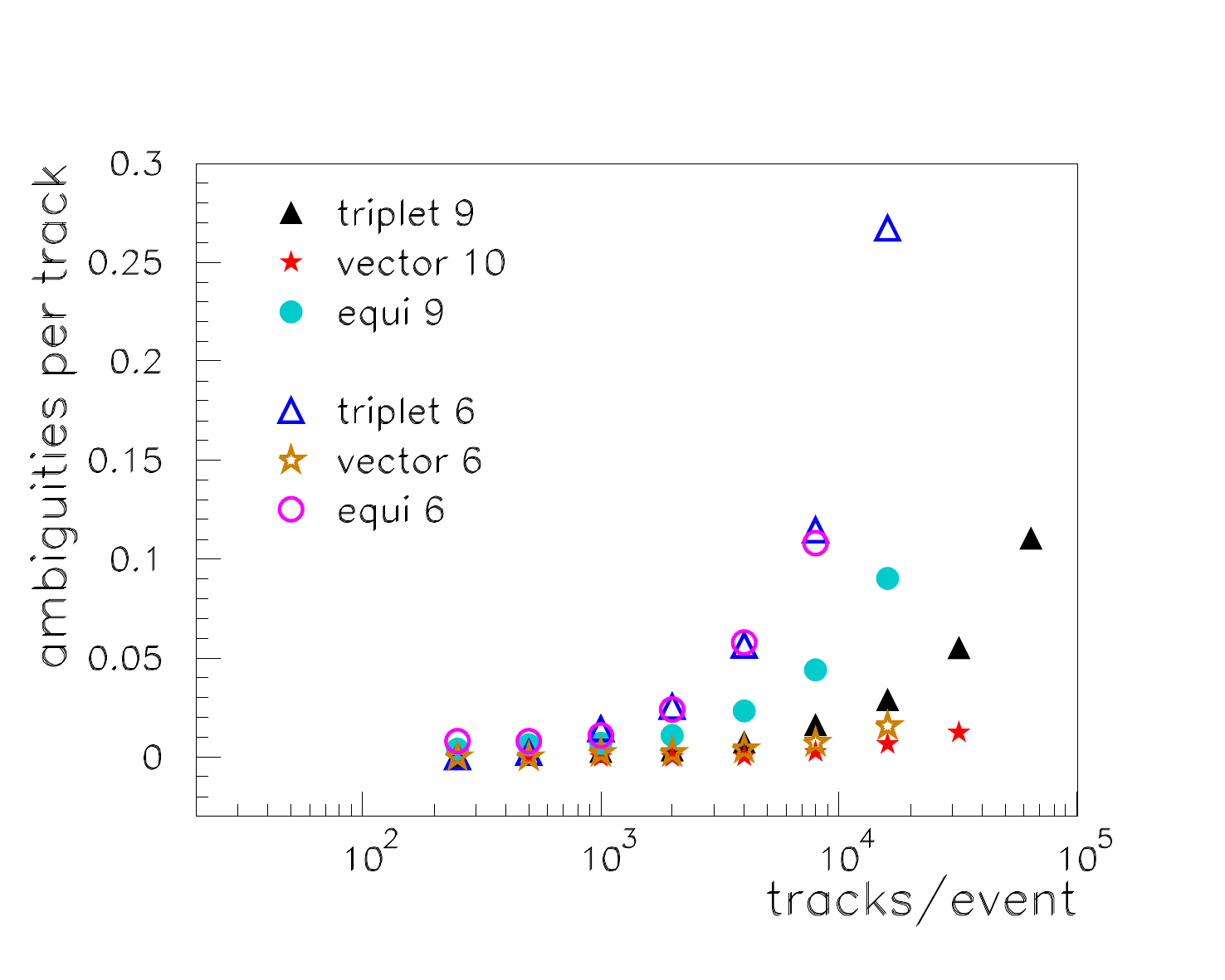}
\put(-14.6,5.5) {(a)}
\put(-7.3,5.5) {(b)}
\vspace{-0.5cm}
\caption{
(a) Simulated number of triplet fits per track as function of the number
of tracks per event. The  {\it dashed} line corresponds to $10^9$ fits per event.
(b)~Simulated fraction of track candidates with hit ambiguities as function of the number
of tracks per event.
The plots show results for the detector geometries defined in 
figure~\protect\ref{fig:6} and for the central detector region $-1.5< \eta < 1.5$.
 }
\end{figure}
The envelopes are shown separately for the transverse and longitudinal plane. 
In the transverse plane the envelope grows quadratically with the
extrapolation distance and reaches about $1$~cm after $25$~cm extrapolation
distance. In contrast,
in the longitudinal plane
 the envelope is only of sub-millimeter size at the same extrapolation distance. 
We can conclude from this study that for efficient track linking
a high $z$-resolution of the detector is much more
important  than a good resolution in the $x$-$y$ plane.~\footnote{
A very good resolution in the $x$-$y$ plane is still required for the
momentum measurement, of course.}
This strongly motivates to use pixel detectors instead of strips!

In a next step we study the track linking performance for the different
geometries of figure~\ref{fig:3}. We use two ``figures of
merit'' to measure the performance: (A) the number of triplet fits needed to validate hit
combinations\footnote{The global track fit of $N$ hits consists of $N-2$
triplet fits.} in the linking procedure and (B) the relative number of tracks
containing hit ambiguities where hits are potentially assigned to more than
one track. 
A link is defined to be valid if a minimum $\chi^2$-cut is fulfilled. The
$\chi^2$-cut is chosen such that $99.5\%$ of the generated tracks are
reconstructed. The simulation and the track fit include multiple scattering
and hit uncertainties.
Figure~\ref{fig:6}~(a) shows the number of triplet fits used for the linking
as function of the
number of tracks per event. Large differences can be seen for the different
detector geometries. 
For the upgraded LHC accelerater about 1000-2000 charged tracks are expected
in the central region.
The ``triplet 9'' design shows an outstanding performance
compared to the other geometries: 
less than 100 triplet fits/track
are required for 1000 tracks/event
whereas for all other geometries significantly more fits are
required.
The ``triplet 9'' geometry is the only one where events with even 64000 tracks
can be completely reconstructed in reasonable time.~\footnote{
Reasonable time is here defined by $10^9$ fits. A single fit on a GPU takes
about $1$~ns \cite{ref:Mu3e_WIT14}.}

\begin{figure}[thb]
\label{fig:7} 
\center
\includegraphics[width=0.5\textwidth] {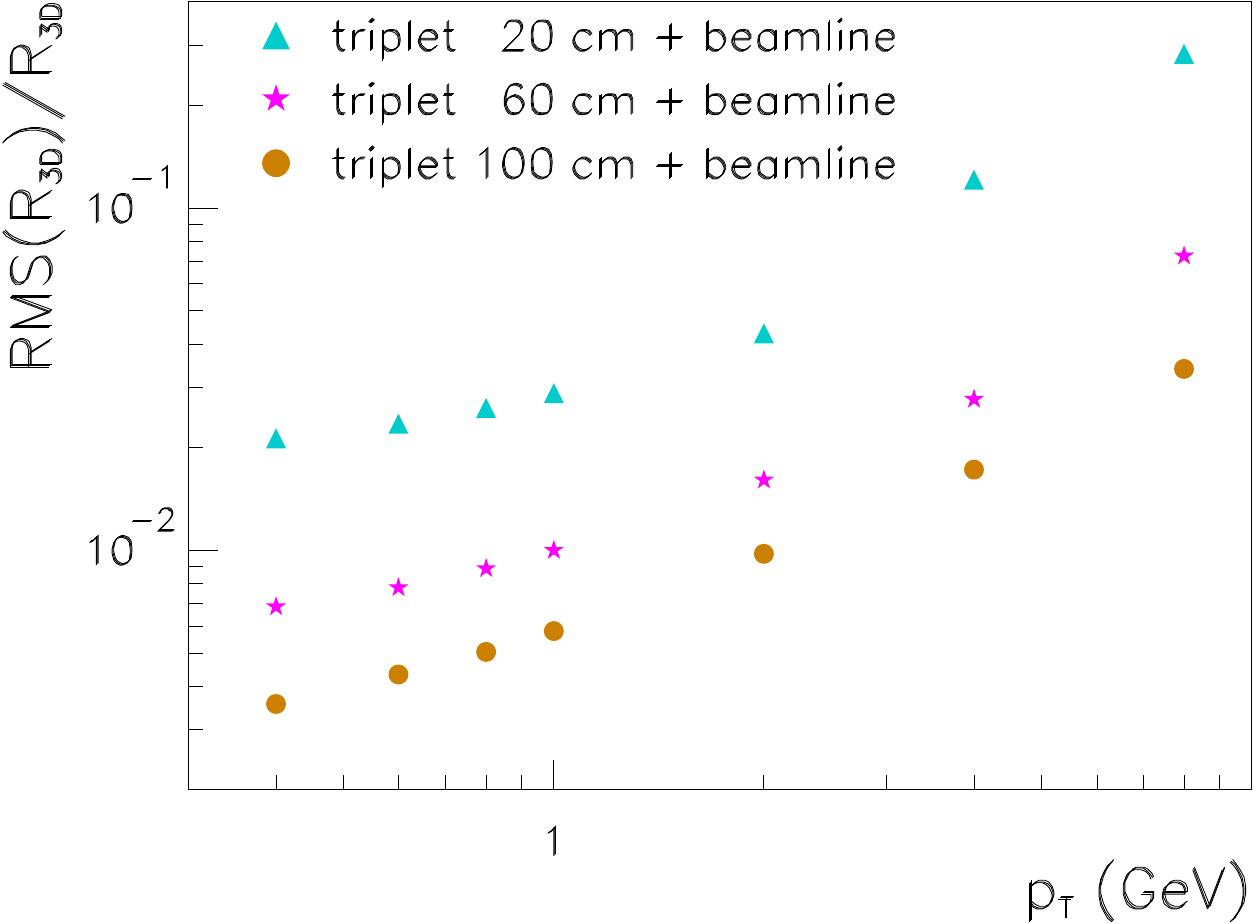}
\caption{
Relative momentum resolution as function of $p_T$
for three different hit triplets with beamline constraint. 
The triplet layers have a radial gap of $\Delta r=1$~cm and are located at radii $r=20$,
$60$ and $100$~cm. 
Simulation parameters are the same as those described in section~\protect\ref{sec:simulation}
%section~\protect{\ref{sec:simulation}}.
}
\end{figure}

Figure~\ref{fig:6}~(b) shows the relative number of tracks with at least one hit
ambiguity, which occurs when a hit can be assigned to more than one track.
Note that a large fraction of the ambiguities could in principal be resolved
using arbitration algorithms. 
However the goal of this study is to quantify the
possible fake rate, and no effort is made to resolve ambiguities here.
We can see that all vector tracker designs and the ``triplet 9'' design have
the lowest rate  of hit ambiguities. 
In general, all designs show a very good
performance for track multiplicities of 1000-2000 per event (LHC-upgrade). 
From these studies we conclude that pixel tracker geometries are very
robust and allow to reconstruct events with very high track multiplicities
at high rate.

\section{Triplet trigger layers}
In the studies discussed so far we have seen that the ``triplet9'' design consisting of three
groups of stacked triplets layers gives overall the
best performance in terms of track linking and track resolution.
Since a single triplet is sufficient to measure track parameters we
study here the simple possibility to use triplet layers for track triggering. 
Compared to designs where sensor layers are spatially separated, 
stacked layers provide the opportunity to correlate hit information between
nearby layers and to form a trigger decision~\cite{ref:garcia2010,ref:Bernardini}.

With a single hit triplet all track parameters can be reconstructed at once and the trajectory
can be extrapolated to the beam-line or the calorimeter. 
In a first step, the very good pointing resolution in the longitudinal plane
and the moderate resolution in the transverse plane (figure~\ref{fig:5})
allows to cross check the primary particle hypothesis that is the track 
originates from the primary vertex.
%The very good pointing resolution in the longitudinal plane together with the moderate
%resolution in the transverse plane (figure~\ref{fig:5}) allows to cross check the
%consistency with originating from the primary vertex in a first step. 
This and the high redundancy in the triplet reconstruction signficantly  reduces 
the fake rate compared to vector tracking designs
where the vertex constraint has to be used for the track reconstruction.
In a second step, the track resolution 
can be further improved by applying the primary vertex constraint. 
The resulting relative momentum resolutions for hit
triplets positioned at radii $r=20,60,100$~cm are shown in figure~\ref{fig:7}.
Tracks with  $p_T=1(10)$~GeV are reconstructed with a resolution of about
1\%(10\%) if the hit triplet is placed at  $r=60$~cm. Such a resolution is
considered to be more than sufficient for a track trigger.

\begin{figure}[htb]
\setlength{\unitlength}{1.0cm}
%\center
\includegraphics[width=0.55\textwidth] {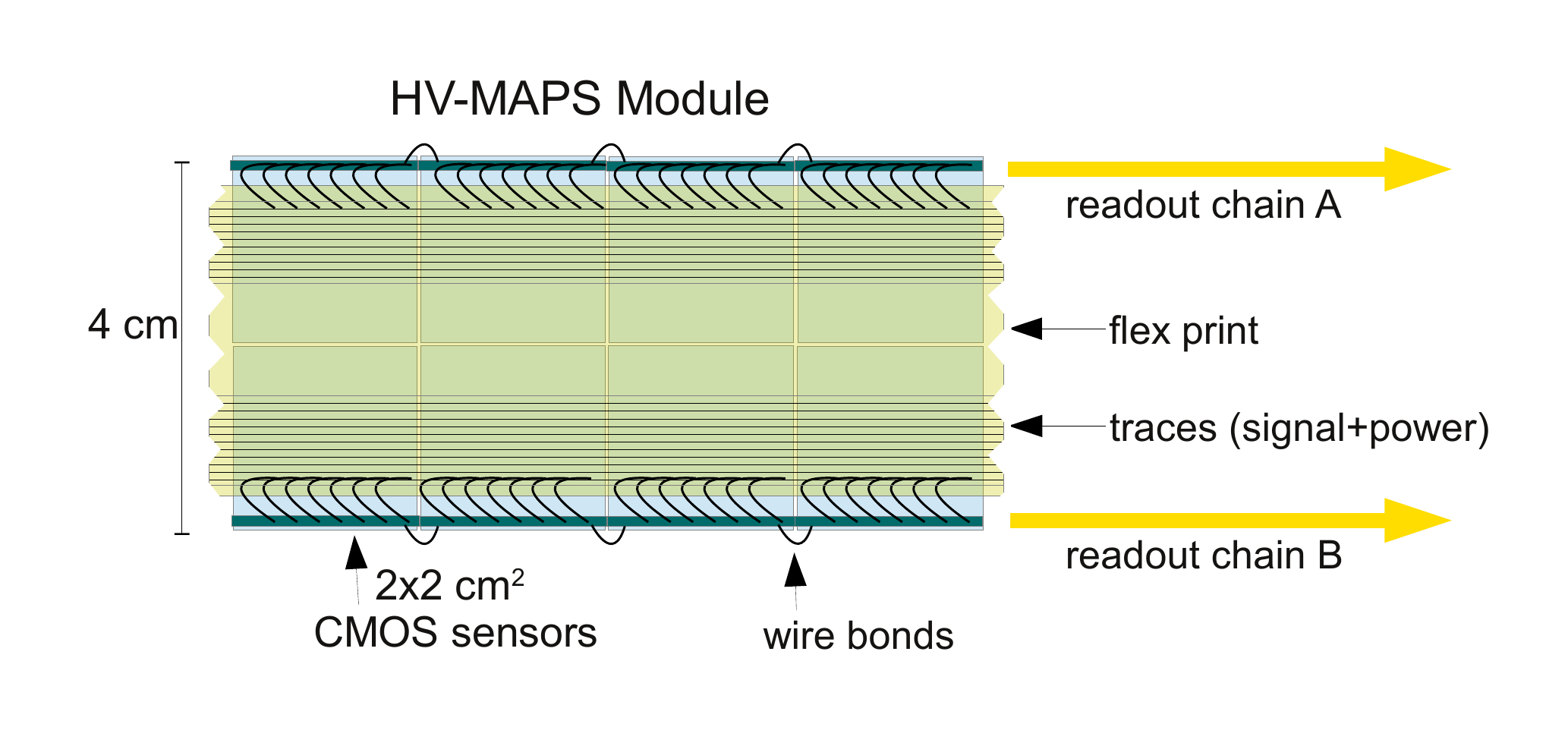}
\put(-7.9,3.6) {(a)}
\includegraphics[width=0.44\textwidth] {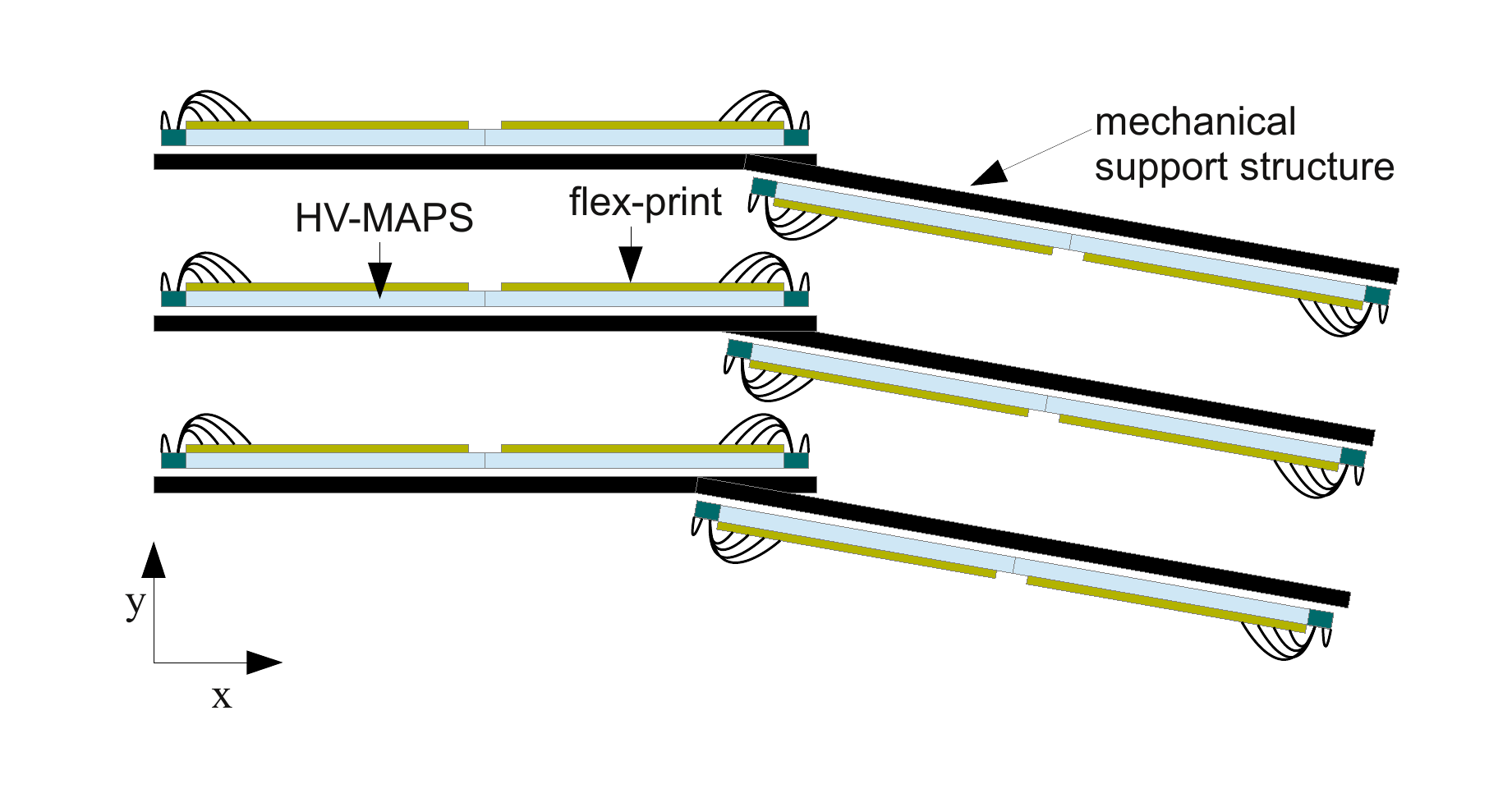}
\put(-5.9,3.6) {(b)}
\caption{(a) Sketch of an ultra-thin pixel ladder consisting of two rows of HV-MAPS reticles
  with a size of $2 \times 2$~cm$^2$. The readout logics and readout drivers
  are located in the small inactive regions at the edges. 
  Electrical connections for signal and powers are made with wire bonds
  to thin flexprints glued on the sensors.
  (b) $x$-$y$ view of three layers of stacked pixel layers. 
  The geometry ensures  high acceptance for track segments solely reconstructed in
  a single trigger tower of three ladders including low momentum tracks.
\label{fig:8} }
\end{figure}

\section{Hardware realisation}
The construction of an all-pixel detector for a LHC-experiment is certainly a big
challenge in terms of radiation hardness, power, cooling, mechanics, and so on. 
Preliminary results, however, indicate that the HV-MAPS
technology might fulfill all these
requirements~\cite{ref:Peric2013,ref:Mu3e_WIT14}, 
and a design
proposal based on this technology including a track trigger option is presented here. 
Figure~\ref{fig:8} shows a section of a $4$~cm wide module composed of $2
\times 2$~cm$^2$ HV-MAPS sensors. About $95\%$ of the sensors surface is
active and small inactive edges contain the readout logic and drivers.
Electrical connections for power, control, monitoring and hit signals are made
via wire bonds. Connections can be done either between adjacent sensors, or between sensors and
multilayer flexprints which are glued directly on the sensors and read out at
one end of the ladder.
Mechanical stability is ensured by gluing the sensors on a carbon-foam
structure which also embeds cooling functionalities.
The readout of a module can be organised by several readout chains. 
Hits are read out after discrimination, digitisation and zero suppression
continuously  via high speed LVDS links, of which several can be
operated in parallel. 

The data rate of a module of $1$~m length
positioned at a radius of $r=80-100$~cm is expected to be about 
$10$~Gbit/s for upgraded LHC luminosities of
${\cal L}=5 \cdot 10^{34} {\rm cm}^{-2}{\rm s}^{-1}$. 
Such data rates could be handled by a fast front-end processor, located at the
end of the ladders, which searches
online for
hit-triplets. 
By using the geometry shown in figure~\ref{fig:8}~(b), hit coincidences between
three stacked layers can be detected with high acceptance, and a trigger
decision signalling for example high momentum tracks can be formed.

\section{Summary}
Various aspects of a new tracking concept based on hit triplets are
discussed for future high rate experiments. 
The focus is put on MS-dominated tracks, relevant for
hadron colliders where most tracks are at low momentum.
At least three layers of pixel detectors are needed for track reconstruction, and
a fast triplet fit based on MS is introduced.
Different types of detector geometries are studied and optimal designs with
emphasis on low momentum tracks are presented.
A design consisting of three groups of triplet layers shows the best reconstruction
performance in terms of speed. 
It is shown that at high particle rates pixel detectors provide a very good track 
reconstruction performance and low fake rate.
Hit triplets can also be used for track triggering, and
a design of an all-pixel tracking detector based on the HV-MAPS
technology is presented.
By using triplet layers in the outer region of a tracking detector,
track triggers could be realised at the upgraded high luminosity LHC experiments.
Such a trigger would be able to reconstruct all tracks
originating from pp interactions for every collision.
More detailed simulations are required to study the trigger rates and 
efficiencies of this concept.

\section{Acknowledgements}
I would like to thank I.~Peric and D.~Wiedner for many inspiring discussions on
the ``all-pixel tracker'' idea and N.~Berger, M. Kiehn and A.~Kozlinskiy for
discussing track reconstruction issues.

\end{document}